\documentclass[twocolumn,showpacs,showkeys, preprintnumbers,prd,superscriptaddress,nofootinbib]{revtex4-1}

\usepackage{amsmath}
\usepackage{amsfonts}
\usepackage{amssymb}
\usepackage{graphicx}
\usepackage{hyperref}
\usepackage{booktabs}
\usepackage{cleveref}
\usepackage{color}

\def\beq{\begin{equation}}
\def\eeq{\end{equation}}

\Crefname{equation}{Eq.}{Eqs.}
\Crefname{figure}{Fig.}{Figs.}
\Crefname{section}{Sec.}{Secs.}

\begin{document}

\title{Black holes and naked singularities from Anton-Schmidt's fluids}

\author{Salvatore Capozziello}
\email{capozzie@na.infn.it}
\affiliation{Dipartimento di Fisica ``Ettore Pancini'', Universit\`a di Napoli ``Federico II'', Via Cinthia, I-80126, Napoli, Italy.}
\affiliation{Istituto Nazionale di Fisica Nucleare (INFN), Sezione di Napoli, Via Cinthia, I-80126, Napoli, Italy.}

\author{Rocco D'Agostino}
\email{rdagostino@na.infn.it}
\affiliation{Dipartimento di Fisica ``Ettore Pancini'', Universit\`a di Napoli ``Federico II'', Via Cinthia, I-80126, Napoli, Italy.}
\affiliation{Istituto Nazionale di Fisica Nucleare (INFN), Sezione di Napoli, Via Cinthia, I-80126, Napoli, Italy.}

\author{Daniele Gregoris}
\email{danielegregoris@libero.it}
\affiliation{Center for Gravitation and Cosmology, College of Physical Science and Technology, Yangzhou University, 180 Siwangting Road, Yangzhou City, Jiangsu Province  225002, China.}
\affiliation{School of Aeronautics and Astronautics, Shanghai Jiao Tong University, Shanghai 200240, China.}

\begin{abstract}
Adopting the Tolman-Oppenheimer-Volkoff formalism, we propose a new analytical solution for a static and spherically symmetric black hole where the cosmological constant is generalized to a non-isotropic Anton-Schmidt fluid acting as a unified dark energy - dark matter source. Our novel result can describe a black hole in energetic equilibrium with the surrounding universe. We thus investigate the interplay between the mass of the black hole and the parameters of the Anton-Schmidt equation of state, and  physically interpret the solution as a family of space-time metrics that may describe both Schwarzschild-de Sitter black holes and naked singularities. The result opens a new window for further constraining the physical properties of the cosmic fluid using the Event Horizon Telescope results, which complement the analysis relying on classical cosmological observations.
\end{abstract}

\date{\today}



\maketitle

\section{Introduction}

The Tolman-Oppenheimer-Volkoff (TOV) approach is a valuable technique to integrate the Einstein field equations of general relativity for a fluid sphere in hydrostatic equilibrium \cite{tov1,tov2,tov3}. Once a polytropic equation of state, relating pressure and energy density of the fluid, is assumed, the same system of differential equations can be applied in the construction of solutions representing gas clouds, stellar configurations, naked singularities, and wormholes \cite{sol1,sol2,sol3,sol4,sol5,sol6,sol7,sol8,sol9,sol10,sol11,sol12}. Also black hole solutions can be understood as limiting-case solutions to the TOV system. The differences between these physical systems are accounted for by the boundary conditions \cite{tov3}: the pressure is zero on the surface of a star, and on the boundary of a gas cloud with the energy density being non-zero and zero in these two cases, respectively. On the other hand, in a black hole solution, a spatial surface, interpreted as the event horizon, must be identified; however, such a surface is not present in the naked singularity solutions. Finally, a throat separating the two space-time sheets must be located in a wormhole metric.

The choice of different equations of state for the matter content entering  Einstein's equations implies different physical properties of the solution. For example, through the mass-radius relation, specific assumptions on the fluids can be tested for neutron stars \cite{ne1,ne2,ne3}, and for white dwarfs \cite{ne4,ne5,ne6}. More in general, wormholes, naked singularities, and black holes can be supported also by exotic (dark) matter-energy. The prototype is the Schwarzschild-de Sitter solution, which involves the presence of a cosmological constant \cite{kottler}. A cosmological constant can be re-interpreted as a dark energy fluid whose pressure and energy density are related by the simple equation of state $p=-\rho$. In the light of  astrophysical datasets, requiring a time-varying equation of state for dark energy \cite{planck}, and for addressing the causality problems  arising from picturing dark energy as a cosmological constant, a number of modified equations of state have been proposed and tested in cosmology: among them,  the Anton-Schmidt scenario represents a relevant example as discussed in \cite{A-S18,A-S19}.

Since there are no theoretical nor observational reasons for assuming that a black hole living inside our Universe does not interact with the surrounding space \cite{large}, it is important to understand how this novel modelling for the dark energy can improve the Schwarzschild-de Sitter solution. Therefore, in this paper we construct a static spherically-symmetric solution of  Einstein's equations for generalizing the former to the case in which the effects of the cosmological constant are mimicked by an Anton-Schmidt fluid. As in \cite{ani1,ani2}, we consider the case of an anisotropic pressure, and we derive an analytical solution supported by a constant energy density. Our solution exhibits a curvature singularity at the center of the manifold. Then, according to specific interplays between the values of the parameters characterizing the Anton-Schmidt fluid, our solution can describe a system of naked singularities in equilibrium, or a Schwarzschild-de Sitter-like black hole with a cosmological and an event horizon. While according to the Cosmic Censorship conjecture, naked singularities should not be created in any realistic description of a star collapse or of a black hole evaporation \cite{pen1,pen2}, they nevertheless exist as solutions of general relativity \cite{naked1}. From the observational point of view, naked singularities and black holes can be distinguished because they accrete matter in different ways \cite{malafarina}. Naked singularities may indeed constitute the final state in processes involving the collisions of black holes \cite{jhep}. Therefore, it is important to clarify under which energy conditions they can theoretically exist. 

Our manuscript is organized as follows: in Sec.~\ref{sII}, we review the basic properties of the Anton-Schmidt fluids and their viability in cosmology. We exhibit the mathematical derivation of the new solution  applying the TOV formalism. In Sec.~\ref{sIII},  the physical interpretation of the solution is discussed, showing that it may represents a system of naked singularities or a generalization of the Schwarzschild-de Sitter solution. Comments on the values of the parameters of the equation of state for discriminating between these two different situations are reported. Finally, we conclude,  in Sec.~\ref{sV}, discussing how this new solution can play a role in the analysis of the geodesic incompleteness of the space-time (in case of naked singularities), or why it could be regarded as a refinement of the existing black hole solutions in the light of the recent Event Horizon Telescope observations \cite{eht, nature}.

\section{Mathematical derivation of the new solutions}\label{sII}

Here, we exhibit the  solution of  Einstein's equations of general relativity for a static spherically-symmetric configuration inside an anisotropic fluid field, whose radial pressure is given by the Anton-Schmidt equation of state. We adopt physical units such that $c=1=G$, and the metric signature $(-,+,+,+)$.

\subsection{The Tolman-Oppenheimer-Volkoff approach}

We start by considering the field equations $G_{\mu\nu}=8\pi T_{\mu\nu}$, where $G_{\mu\nu}$ is the Einstein tensor. Assuming to deal with a perfect fluid with an anisotropic pressure, the stress-energy tensor takes the form $T^{\mu}{}_{\nu}={\rm diag}(-\rho,\, p_r,\,p_t,\,p_t)$, where $\rho$ is the fluid energy density, $p_r$ its radial pressure, and $p_t$ its transverse pressure. 
We can then write the metric for a static, spherically symmetric configuration in the system of coordinates ($t$, $r$, $\theta$, $\phi$) as \cite{exact}
\beq
\label{metric1}
ds^2=-e^{2\Phi(r)} dt^2 + e^{2 \alpha(r)} dr^2 + r^2 d\Omega^2 .
\eeq
In our scheme, the energy density and pressure depend only on the radial coordinate, namely $\rho=\rho(r)$, $p_r=p_r(r)$ and $p_t=p_t(r)$. The coefficient of anisotropy of the pressure is defined as $a(r)=(p_r -p_t)/r$.  $\Phi(r)$ is known as the redshift function \cite{klein}, while the function $b(r)=r(1-e^{-2\alpha(r)})$ is the shape function. The mass function of the configuration is $M(r)=\frac{r}{2}(1-e^{-2\alpha(r)})$ \cite{morris}.
The corresponding Einstein's equations for the metric (\ref{metric1}) are
\begin{eqnarray}
\label{eqrho1}
&& 2 \frac{d \alpha(r)}{dr} r +e^{2\alpha(r)} -1 \,=\,  8 \pi \rho(r) r^2 e^{2 \alpha(r)}\ , \\
\label{eqpr}
&& 2 \frac{d \Phi(r)}{dr} r -e^{2\alpha(r)} +1 \,=\,  8 \pi p_r(r) r^2 e^{2 \alpha(r)}\ , \\
\label{eqpt}
&& \frac{d^2 \Phi(r)}{dr^2} r +\left(\frac{d \Phi(r)}{dr} -  \frac{d \alpha(r)}{dr}  \right)\left( r \frac{d \Phi(r)}{dr} +1\right) \nonumber\\ &&\,=\, 8 \pi e^{2 \alpha(r)} r p_t(r)\,.
\end{eqnarray}
In general, for a configuration in hydrodynamical equilibrium, this system of differential equations must be complemented with the contracted Bianchi identities $T^{\mu\nu}{}_{;\nu}=0$, where the semicolon denotes covariant derivative. In this case, the only non-zero component is $\mu=r$, which reads
\beq
(\rho(r)+p_r(r)) \frac{d \Phi(r)}{dr}+\frac{d p_r(r)}{dr}+a(r)=0 \ .
\eeq
This equation is automatically satisfied once Eqs.~(\ref{eqrho1})-(\ref{eqpt}) are assumed to hold. We note that system (\ref{eqrho1})-(\ref{eqpt}) represents the anisotropic generalisation of the well-known Tolman-Oppenheimer-Volkoff (TOV) equations, which have been extensively used throughout the literature for the construction of stellar configurations in hydrodynamic equilibrium \cite{tov1,tov2,tov3} (see also Tables 16.1 and 16.2 of \cite{exact} for a list of analytical solutions).

\subsection{The Anton-Schmidt fluids}

The novelty of our approach consists in assuming that the radial pressure obeys the Anton-Schmidt equation of state \cite{A-S18,A-S19,Boshkayev19}:
\beq
p_r(\rho)= A\left(\frac{\rho}{\rho_*}  \right)^{-n} \ln \frac{\rho}{\rho_*} \,.
\label{eq:Anton-Schmidt}
\eeq
The Anton-Schmidt scenario has been introduced in the cosmological context to unify dark matter and dark energy into a single fluid \cite{A-S18,Odinstov18}. Analogously to the Chaplygin gas \cite{Chaplygin}, it is possible to consider the cosmic fluid made of matter characterized by a non-vanishing equation of state, which would account for the accelerated expansion observed at late epochs of the universe evolution. In the original proposal, the Anton-Schmidt pressure was invoked to describe the deformation of crystalline solids under isotropic stress within the  Debye approximation \cite{Anton97}. In a similar way, considering the action of the cosmic expansion which makes the universe deform, one can model the evolution of the cosmic fluid through an equation of state that becomes naturally negative at certain epochs. 

In \Cref{eq:Anton-Schmidt}, the constant $A$ is linked to the bulk modulus of a crystal at the equilibrium volume, and $n$ is related to the Gr\"uneisen parameter proper of the material (see \cite{A-S18} for more details); $\rho_*$ represents a reference density that has been interpreted as the Planck density in \cite{Chavanis15}, based on observational considerations on the surface density profile of dark matter halos resulting from the study of the logotropic equation of state. In this respect, the Anton-Schmidt picture can be seen as a generalization of the logotropic fluid, which is recovered in the limit $n=0$.
Interestinlgy, the Anton-Schmidt model with $n=0$ was found to match the Lagrangian of the extended Chaplygin gas in non-relativistic regime \cite{Boshkayev19}.

An Anton-Schmidt fluid is characterized by the following adiabatic sound speed:
\beq
c^2_s\equiv\frac{\partial p_r}{\partial \rho}=\frac{A}{\rho}\left(\frac{\rho}{\rho_*}  \right)^{-n}-\frac{n p_r}{\rho}\ .
\label{eq:c_s}
\eeq
This can be integrated to obtain the corresponding equation of state parameter:
\begin{equation}
w\equiv \dfrac{p_r}{\rho}=A\left(\dfrac{\rho}{\rho_*}\right)^{-n}\ln\left(\dfrac{\rho}{\rho_*}\right) +\dfrac{C}{\rho}\ ,
\end{equation}
where the constant of integration $C$ is usually set to zero.
In the Anton-Schmidt paradigm, dark matter and dark energy are unified into a single fluid that encodes different cosmic behaviours. 
This reveals as a concrete alternative to the standard model, where the effects of the late-time acceleration are accounted for through the \emph{ad hoc} introduction of the cosmological constant with a negative equation of state. As shown in \cite{A-S18}, in the early universe dominated by the rest-mass energy, the Anton-Schmidt pressure tends to vanish consistently with a matter-dominated universe scenario. After a transition epoch at which the universe's volume $(V\propto \rho^{-1})$ is equal to the equilibrium volume, the pressure tends to a negative constant value for $n < 0$, mimicking the late-time effects of the cosmological constant.

\subsection{Integrating the TOV equations for a constant energy density}

In order to integrate the TOV equations under the Anton-Schmidt equation of state hypothesis, we take advantage of the Lambert special function $W(x)$, satisfying the relation $W(x)e^{W(x)}=x$ and the following identity  \cite{lambert1,lambert2}:
\beq
e^{l W(x)} = \left(\frac{x}{W(x)}   \right)^l   .
\eeq
We can then invert (\ref{eqpr}) into 
\beq
\rho(r)= \rho_* \left[\frac{ W(\chi)}{\chi}   \right]^{\frac{1}{n}}\,, \quad \chi=\frac{1-e^{-2 \alpha(r)}\left( 2r \frac{d \Phi(r)}{dr}+1  \right)}{8 \pi A r^2}\,,
\eeq
In the particular case \footnote{We will show in the next section that this choice corresponds to have a constant energy density $\rho(r)=\frac{3A}{8\pi}$.} $\chi=e$, it is possible to construct an analytical solution for the TOV equations in terms of elementary functions.  In fact, one obtains
\beq
\label{eqphi}
\Phi(r)= \int \frac{n(e^{2 \alpha(r)} -1) - 8 \pi A r^2 e^{2\alpha(r)+1}}{2nr}dr +{\mathcal C}_1\ ,
\eeq
where ${\mathcal C}_1$ is an arbitrary constant of integration. Furthermore, Eq.~(\ref{eqrho1}) can be integrated into
\beq
\label{alphaexp}
\alpha(r)=-\frac{1}{2} \ln\left(1+\frac{{\mathcal C}_2}{r}-\frac{8 \pi \rho_* e^{-1/n} }{3} r^2  \right) ,
\eeq
where ${\mathcal C}_2$ is another constant of integration. We note that, for the particular choice ${\mathcal C}_2=0$,  the case of the interior Schwarzschild solution is reproduced  \cite{exact}, and that the metric coefficient $g_{rr}$ is asymptotically flat. Hence,  one can rewrite Eq.~(\ref{eqphi}) as
\beq
\Phi(r)=\frac{1}{2n}\int \frac{(24 \pi A e -8 \pi n \rho_* e^{-1/n}) r^3 +3 {\mathcal C}_2 n}{(8 \pi \rho_* e^{-1/n} r^3 -3  {\mathcal C}_2 -3r )r}dr+ {\mathcal C}_1 \,.
\eeq
Now, let $R_i$ ($i=1,2,3$) be the three roots of the algebraic equation
\beq
\label{cubic}
8 \pi \rho_* e^{-1/n} R^3 -3R -3 {\mathcal C}_2=0\,,
\eeq
whose discriminant reads \cite{lambert2}
\beq
\Delta=216 \pi \rho_* e^{-1/n}(4-72\pi \rho_* e^{-1/n}{\mathcal C}_2^2)\,.
\eeq
When $\Delta>0$, i.e. when the reference energy density satisfies 
\beq
\label{3real}
0<\rho_*<\frac{e^{1/n}}{18 \pi {\mathcal C}_2^2}\,,
\eeq
there are three real distinct roots; for $\Delta=0$,  which corresponds to
\beq
\label{repeat}
\rho_* =\frac{e^{1/n}}{18 \pi {\mathcal C}_2^2}\,,
\eeq
there are three real roots with one of them of multiplicity two; instead for $\Delta<0$, that is
\beq
\rho_* >\frac{e^{1/n}}{18 \pi {\mathcal C}_2^2}\,,
\eeq
one root is real and two are complex conjugate.  Explicitly, the three roots of Eq.~(\ref{cubic}) are
\begin{align}
&R_1 = S_1 +S_2\ , \\
&R_{2,3}= -\frac{S_1+S_2}{2}\pm \frac{i \sqrt{3}}{2}(S_1 -S_2) \ ,
\end{align}
where
\begin{equation}
    S_{1,2} =  \left[\frac{ e^{1/n}}{8 \pi \rho_*}  \left(\frac{3 {\mathcal C}_2}{2} \pm \sqrt{\frac{9 {\mathcal C}_2^2}{4} - \frac{ e^{1/n}}{8 \pi \rho_*}} \right)\right]^{1/3}  .
\end{equation}
Therefore, for the redshift function, we finally get
\begin{equation}
\label{phiexp}
\Phi(r) = \frac{1}{2}\left( \sum_{i=1}^{3} \ln (r-R_i)^{\gamma_i}  - \ln r \right)+{\mathcal C}_1\ ,
\end{equation}
where 
\begin{equation}
\gamma_i = \frac{n-8\pi A eR^2_i}{n(1- 8\pi \rho_* e^{-1/n} R_i^2)} \ .
\end{equation}
Since a non-zero ${\mathcal C}_1$ corresponds to a scaling of the time variable $t$, which might be re-absorbed through a coordinate transformation, we can set ${\mathcal C}_1=0$ without loss of generality.  Thus,
\beq
\label{gtt}
g_{tt}(r)=-\frac{1}{r} \prod_{i=1}^{1=3} (r-R_i)^{\gamma_i}.
\eeq

\section{Physical interpretation of the new solutions} \label{sIII}

In this section, we  provide the physical interpretation of the mathematical solutions we have derived. Let us distinguish between the cases of a stellar configuration, a gas cloud, a wormhole, a black hole, and a naked singularity. Our analysis allows us to illuminate the role of the parameters entering the Anton-Schmidt equation of state in the context of hydrodynamic equilibrium with anisotropic pressure. Our classification concerns with the boundary conditions that can be associated to the TOV differential equations.
\\

A star does not have any physical singularity. However, we note that Eq.~(\ref{gtt}) seems to have a pole in $r=0$. Since $R=0$ cannot be a solution  of Eq.~(\ref{cubic}), unless in the trivial case $\mathcal C_2=0$ that we have already discussed in the previous section, this pole cannot be removed. Therefore, our solution exhibits a curvature singularity in $r=0$ and cannot be interpreted as modelling a stellar configuration.

A gas cloud interpretation requires to provide the boundary of the cloud (possibly located also at infinity) at which both the energy density and the radial pressure vanish. However, in our solution,  the energy density is spatially homogeneous and then, if it vanishes in one spatial point, it vanishes everywhere. In this latter case, we would have a vacuum solution (in fact also the radial pressure is vanishing when the energy density is vanishing, i.e. for $A=0$), and not a gas cloud.

According to \cite{morris}, we can physically interpret the mathematical solution of the TOV equations derived in the previous section as describing a wormhole if the following conditions are met:
\begin{eqnarray}
&&\lim_{r \to r_0^+} \alpha(r)=+\infty\ , \quad b(r_0)=r_0\ , \\
&& \frac{d b(r)}{dr}_{r=r_0}<1\,, \quad b(r)<r\ .
\end{eqnarray}
In addition, $e^{2 \Phi(r)}$ must be finite and non-vanishing in the region near the throat of the wormhole, which is located at $r_0$, and the coordinate system is such that $r>r_0$.
 From Eq.~(\ref{alphaexp}), specifically we get
 \beq
 \label{C2a}
 {\mathcal C}_2=r_0 \left(   \frac{8 \pi \rho_* e^{-1/n}}{3}r_0^2 -1 \right).
 \eeq
 Then, the condition $b(r)<r$ can be recast into
 \beq
 r\left( \frac{8 \pi \rho_* e^{-1/n}}{3} r^2 -1\right)< r_0\left( \frac{8 \pi \rho_* e^{-1/n}}{3} r_0^2 -1\right).
 \eeq
 which clearly cannot hold for $r>r_0$. Therefore, our solution is qualitatively different from \cite{luis},  which proposed a wormhole solution supported by a constant energy density with anisotropic pressure.

 \subsection{Black hole vs. naked singularity}
 
The Schwarzschild-de Sitter solution reads \cite{exact, kottler}
 \begin{equation}
 ds^2= -f(r)dt^2+\frac{dr^2}{f(r)}+r^2(d\theta^2+\sin^2\theta d\phi^2)\ ,
 \end{equation}
 where
 \begin{equation}
 f(r)=1-\frac{2M}{r}-\frac{\Lambda r^2}{3}\ , 
 \end{equation}
 being $M$ the mass of the black hole and $\Lambda$ the cosmological constant. Comparing with Eq.~(\ref{alphaexp}), it is natural to interpret
 \begin{eqnarray}
 \label{C2}
{\mathcal C}_2 &=& -2M ,\\
\Lambda_\text{eff}&=& 8\pi \rho_* e^{-1/n}>0 \,, \nonumber
 \end{eqnarray}
 where $\Lambda_\text{eff}$ plays the role of an effective cosmological constant. The redshift function in the Schwarzschild-de Sitter solution can be rewritten as
 \beq
 \label{gttk}
 g_{tt}=-\frac{3r -6M -\Lambda r^3}{3 r}\,,
 \eeq
 showing a pole in $r=0$, which appears as well in our new solution Eq.~(\ref{gtt}). The horizons are located in the zeros of Eq.~(\ref{gttk}). We note that, for $\Lambda>0$, the cubic equation
 \beq
 \label{hkot}
 \Lambda r^3 -3r +6M =0
 \eeq
admits three real roots if $1-9 \Lambda M^2\geq0$. The product of the three roots must be equal to $-6M/\Lambda<0$ \cite{lambert2}, and therefore we can have two cases: 
\begin{enumerate}
\item Three negative roots: in this case,  the solution has no horizon.
\item One negative root and two positive roots, which can can be interpreted as the black hole event horizon and the inflationary cosmological horizon in the middle of which the observer is placed.
\end{enumerate}

\noindent Now, Eq.~(\ref{cubic}), together with  Eq.~(\ref{C2}), constitute the generalization of Eq.~(\ref{hkot}). Fist of all, we note that the condition for having three real roots become
\beq
0<\rho_* \leq \frac{e^{1/n}}{72 \pi M^2}\,.
\eeq
Moreover, in the case of three real negative roots, our solution describes a naked singularity located in $r=0$, like in the case of the Schwarzschild-de Sitter solution. However, in the case in which we have one negative and two positive real roots, the situation can be qualitatively different than the one met for the Schwarzschild-de Sitter black hole. In fact, the exponents in Eq.~(\ref{gtt}), in principle, can be either positive or negative according to certain combinations of the numerical values for the parameters entering the Anton-Schmidt equation of state. When they are positive, we have horizons, while, when they are negative, we have singularities. Therefore, our solution may be interpreted as being characterized by three naked singularities, two singularities and one horizon, or one singularity and two horizons.

\section{Discussion and Conclusions}\label{sV}

Supermassive black holes are expected to exchange energy with their surrounding cosmic environment  because their growth is intimately related to the evolution of their host galaxy. Since these effects still occur in the present epoch of the Universe, the possibility that dark matter and dark energy may be absorbed onto the black must be accounted for \cite{astror1,astror2,astror3,astror4}. To a first approximation, one can assume that the cosmic fluid evolves according to the Friedmann cosmology, and that the rate of increase/decrease of the mass of the black hole is directly proportional to the area of its surface and to the sum of energy and pressure of the dark fluid \cite{ev1,ev2,ev3,ev4}. An improved treatment couples the evolution of the mass of the black hole to the evolution of the Universe adopting the McVittie metric, which does not require any \emph{ad hoc} assumptions on the rate of energy exchange that is automatically implemented through the Einstein field equations \cite{mcvittie}.

In this paper, we have proposed a  solution describing a black hole that has reached an equilibrium with its environment and no further energy exchanges occur. The role of dark energy and dark matter are both accounted for through the Anton-Schmidt fluid, which provides an elegant and complete picture of their unification. Therefore, applying the TOV formalism in general relativity, we constructed an exact and analytical metric for a static and spherically symmetric black hole that generalizes the  Schwarzschild-(a)deSitter space-time by replacing the cosmological constant with a more general dark fluid and allowing an anisotropic pressure. The former improvement is needed in light of the outcomes of the Planck mission, which favor an evolving dark energy equation of state over the simplest scenario of a cosmological constant \cite{planck}.

By investigating the role of the free parameters entering our new solution, we could claim that it actually represents a family of space-time metrics describing both black holes and naked singularities depending on certain relations involving the mass of the black hole and the parameters of the equation of state. Thus, in light of the Censorship Conjecture, we were able to provide a novel theoretical constraint between the cosmological parameters of the Anton-Schmidt scenario. Such restrictions may, in principle, be further sharpened by comparing the estimated size of the black hole to the recent observations provided by the Event Horizon Telescope \cite{eht, nature}, and combined with the previous analyses of the Anton-Schmidt fluid dealing with low-redshift cosmological observables and the cosmic microwave background measurements.

\section*{Acknowledgements}
S.C. and R.D. acknowledge the support of INFN (iniziative specifiche MoonLIGHT-2 and QGSKY).
D.G. acknowledges the support from China Postdoctoral Science Foundation (grant No. 2019M661944).

{}

\end{document}